\documentclass[a4paper,aps,preprint]{revtex4}

\usepackage{amsfonts, amsbsy, amssymb, amsmath, graphicx, float}

\newcommand{\bx}{\boldsymbol{x}}

\begin{document}

\title{Bulgac-Kusnezov-Nos\'e-Hoover thermostats}

\author{Alessandro Sergi}
\email{sergi@ukzn.ac.za}

\affiliation{
School of Physics, University of KwaZulu-Natal, Pietermaritzburg,
Private Bag X01 Scottsville, 3209 Pietermaritzburg, South Africa}

\author{Gregory S. Ezra}
\email{gse1@cornell.edu}
\affiliation{Department of Chemistry and Chemical Biology,
Baker Laboratory, Cornell University, Ithaca, New York 14853, USA}

\begin{abstract}
In this paper we formulate Bulgac-Kusnezov constant temperature dynamics
in phase space by means of non-Hamiltonian brackets.
Two generalized versions of the dynamics are similarly defined:
one where the Bulgac-Kusnezov demons are globally controlled
by means of a single additional Nos\'e variable,
and another where each demon is coupled to
an independent Nos\'e-Hoover thermostat.
Numerically stable and efficient
measure-preserving time-reversible algorithms are derived
in a systematic way for each case.
The chaotic properties of the different phase space flows
are numerically illustrated
through the paradigmatic example of the one-dimensional harmonic oscillator.
It is found that, while the simple Bulgac-Kusnezov thermostat
is apparently not ergodic, both of
the Nos\'e-Hoover controlled dynamics sample the
canonical distribution correctly.
\end{abstract}

\today

\maketitle

\section{Introduction}

In condensed matter studies, there are many situations
in which molecular dynamics simulation
at constant-temperature~\cite{Nose91,Frenkel01,Leimkuhler04a} is needed.
For example, this occurs when magnetic systems are modelled in
terms of classical spins~\cite{Chen94,Bunker96,Evertz96,Costa97}.
Deterministic methods \cite{Nose84,Hoover85,Jellinek88a},
based on non-Hamiltonian dynamics~\cite{Evans90,Tuckerman99,Tuckerman01,Sergi01,Sergi03,Sergi07,Ezra02,Ezra04,Ezra06},
can sample the canonical distribution
provided that the motion in the phase space of
the relevant degrees of freedom is ergodic  \cite{Nose91,Leimkuhler04a}.
However, classical spin systems are usually formulated in terms of
non-canonical variables \cite{Bulgac90a,Marsden99}, without a
kinetic energy expressed through momenta in phase space,
so that Nos\'e dynamics cannot be applied directly.
To tackle this problem, Bulgac and Kusnezov (BK) introduced
a  deterministic constant-temperature dynamics~\cite{Bulgac90,Kusnezov90,Kusnezov92}
which can be applied to spins.
A number of numerical approaches to integration of spin dynamics
can be found in the literature~\cite{Frank97,Arponen04,Tsai04,McLachlan06}.
However, BK dynamics, as any other deterministic canonical
phase space flow, is able to correctly sample the canonical
distribution \emph{only} if the motion in phase space is ergodic on the timescale of the simulation.
In general, this condition is very difficult to check for statistical
systems with many degrees of freedom, while it is known
that, despite its simplicity, the one-dimensional harmonic oscillator provides
a difficult and important challenge for deterministic thermostatting methods
\cite{Hoover85,Hoover01a,Legoll07,Legoll09}.

In this paper, we accomplish two goals.
First, by reformulating BK dynamics through non-Hamiltonian
brackets~\cite{Sergi01,Sergi03} in phase space,
we introduce two generalized versions
of the BK time evolution
which are able to sample the canonical distribution for a stiff harmonic
system.
Second, using a recently introduced approach based on the geometry
of non-Hamiltonian phase space~\cite{Ezra06},
we are able to derive stable and efficient
measure-preserving and time-reversible
algorithms in a systematic way for all the phase space flows treated here.

The BK phase space flow introduces temperature control by means
of fictitious coordinates (and their associated momenta in an
extended phase space)
traditionally called `demons'.
Our generalizations of the BK dynamics are obtained by controlling
the BK demons themselves by means of additional Nos\'e-type variables \cite{Nose84}.
In one case, the BK demons  are controlled  globally by means of
a single additional Nos\'e-Hoover thermostat~\cite{Nose84,Hoover85}.
In the following this will be referred to as (BKNH) Bulgac-Kusnezov-Nos\'e-Hoover dynamics.
In the second case, each demon is coupled to an independent
Nos\'e-Hoover thermostat.
This will be called the Bulgac-Kusnezov-Nos\'e-Hoover chain (BKNHC), and
corresponds to  `massive' NH thermostatting of the demon variables \cite{Martyna92}.

The ability to derive numerically stable
measure-preserving time-reversible
algorithms \cite{Ezra06} for Nos\'e controlled BK dynamics is very encouranging
for future applications to thermostatted spin systems.

This paper is organized as follows.
In Sec.~\ref{sec:nh-bra} we briefly sketch the unified formalism for
non-Hamiltonian phase space flows and measure-preserving integration.
The BK dynamics is formulated in phase space and a measure-preserving
integration algorithm is derived in Sec.~\ref{sec:bk}.
The BKNH and BKNH-chain thermostats are treated in Secs.~\ref{sec:cbk}
and~\ref{sec:nbkc} respectively.
Numerical results for the one-dimensional harmonic oscillator using these
thermostats
are presented and discussed in Sec.~\ref{sec:num}.
Section~\ref{sec:concl} reports our conclusions.

In addition we include several Appendices.
A useful operator formula is derived in Appendix~\ref{app:app1},
while invariant measures for the BK, BKNH, and BKNHC phase space flows
are derived in Appendices \ref{app:BK-stat},~\ref{app:CBK-stat},
and~\ref{app:NBK-stat}, respectively.

\newpage

\section{Non-Hamiltonian brackets and measure-preserving algorithms}
\label{sec:nh-bra}

Consider an arbitrary system admitting a time-independent (extended) Hamiltonian
expressed in terms of the phase space coordinates $x_i$, $i=1,\ldots, 2N$.
In this case, the Hamiltonian can be interpreted as
the conserved energy of the system.

Upon introducing an antisymmetric tensor field (generalized Poisson tensor \cite{Ramshaw91,Marsden99}) in phase space,
$\mbox{\boldmath$\cal B$}(\bx)=-\mbox{\boldmath$\cal B$}^T(\bx)$,
one can define non-Hamiltonian brackets~\cite{Sergi01,Sergi03,Sergi07} as
\begin{equation}
\left\{a,b\right\}=\sum_{i,j=1}^{2n}\frac{\partial a}{\partial x_i}
{\cal B}_{ij}\frac{\partial b}{\partial x_j}
\;,\label{eq:nh-bra}
\end{equation}
where $a=a(\bx)$ and $b=b(\bx)$ are two arbitrary
phase space functions.
The bracket defined in Eq.~(\ref{eq:nh-bra}) is classified as
non-Hamiltonian~\cite{Sergi01,Sergi03,Sergi07} since, in general, it does not obey
the Jacobi relation, i.e., in general the Jacobiator ${\cal J} \ne 0$, where 
\cite{Marsden99}
\begin{equation}
{\cal J}=\left\{a,\left\{b,c\right\}\right\}
+\left\{b,\left\{c,a\right\}\right\}
+\left\{c,\left\{a,b\right\}\right\}\;,
\end{equation}
with $c=c(\bx)$ arbitrary phase space function (in addition to
the functions $a$ and $b$, previously introduced).
If ${\cal J} \neq 0$, the tensor
${\cal B}_{ij}$ is said to define an `almost-Poisson' structure \cite{DaSilva99}.
(Such systems have also been called `pseudo-Hamiltonian' \cite{Ramshaw91}.)

An  energy-conserving and in general non-Hamiltonian phase space flow is then defined
by the vector field
\begin{equation}
\dot{x}_i=\left\{x_i,H\right\}=\sum_{j=1}^{2N}{\cal B}_{ij}
\frac{\partial H}{\partial x_j}\;,
\label{eq:gen-psf}
\end{equation}
where conservation of $H(\bx)$ follows directly from the
antisymmetry of ${\cal B}_{ij}$.

It has previously been shown how equilibrium statistical mechanics can be comprehensively
formulated within this framework~\cite{Sergi07}.
It is also possible to recast the above formalism and the corresponding statistical mechanics
in the language of differential forms~\cite{Ezra02,Ezra04}.
If the matrix ${\cal B}$ is invertible
(this is true for all the cases considered here), with inverse $\Omega_{ij}$,
we can define the 2-form \cite{Schutz80}
\begin{equation}
\label{eq:2-form}
\Omega = \tfrac12 \Omega_{ij} d x^i \wedge  d x^j.
\end{equation}
The dynamics of Eq.~(\ref{eq:gen-psf}) is then Hamiltonian if and only if the form (\ref{eq:2-form})
is \emph{closed}, i.e., has zero exterior derivative, $d \Omega = 0$ \cite{Schutz80}.
This condition is independent of the particular system of coordinates used
to describe the dynamics.

The structure of Eq.~(\ref{eq:gen-psf})
can be taken as the starting point for derivation of efficient
time-reversible integration algorithms that also preserve
the appropriate measure in phase space~\cite{Ezra06}.
Measure-preserving algorithms can be derived upon introducing a splitting
of the Hamiltonian
\begin{equation}
H=\sum_{\alpha=1}^{n_s}H_{\alpha}
\label{eq:Hdecomp}
\end{equation}
which in turn induces a splitting of the Liouville operator
associated with the non-Hamiltonian bracket in Eq.~(\ref{eq:nh-bra}),
\begin{equation}
L_{\alpha}x_i=\left\{x_i,H_{\alpha}\right\}=\sum_{j=1}^{2N}{\cal B}_{ij}
\frac{\partial H_{\alpha}}{\partial x_j}\;.
\label{eq:Ldecomp}
\end{equation}
When the phase space flow has a non-zero compressibility
\begin{equation}
\kappa=\sum_{i,j=1}^{2N}\frac{\partial{\cal B}_{ij}}{\partial x_i}
\frac{\partial H}{\partial x_j}
\end{equation}
the statistical mechanics must be formulated in terms of a modified phase space measure
\cite{Tuckerman99,Tuckerman01,Sergi01,Sergi03,Sergi07,Ezra02,Ezra04}
\begin{equation}
\overline{\omega}=e^{-w(x)}\omega
\end{equation}
where
\begin{equation}
\omega=dx^1 \wedge dx^2 \wedge \ldots \wedge dx^{2N}
\end{equation}
is the standard phase space volume element (volume form \cite{Schutz80})
and the statistical weight $w(x)$ is defined by
\begin{equation}
\frac{d w }{dt}=\kappa(x)\;.
\end{equation}
It has been shown that, provided the condition
\begin{equation}
\label{eq:div}
\frac{\partial}{\partial x_j}  \;\left[ e^{-w(x)} {\cal B}_{ij}\right] = 0, \;\; i = 1, \ldots 2N
\end{equation}
is satisfied, then
\begin{equation}
L_{\alpha}\overline{\omega}=0\quad {\rm for~every}~\alpha\;,
\end{equation}
so that the volume element $\overline{\omega}$ is invariant under
each of the $L_\alpha$~\cite{Ezra06}.
The condition (\ref{eq:div}) is satisfied for all the cases considered below,
so that, exploiting the decomposition in Eq.~(\ref{eq:Ldecomp}),
algorithms derived by means
of a symmetric Trotter factorization of the Liouville propagator:
\begin{eqnarray}
\exp[\tau L]&=&
\prod_{\alpha=1}^{n_s-1}\exp\left[\frac{\tau}{2}L_{\alpha}\right]
\exp\exp\left[\tau L_{n_x}\right]
\nonumber\\
&\times&
\prod_{\beta=1}^{n_s-1}\exp\left[\frac{\tau}{2}L_{n_s-\beta}\right]
\end{eqnarray}
are not only time-reversible but also measure-preserving.

\newpage

\section{Phase space formulation of the BK thermostat}
\label{sec:bk}

A phase space formulation of the BK thermostat can be achieved
upon introducing the Hamiltonian
\begin{subequations}
\begin{align}
H^{\rm BK}
&=\frac{p^2}{2m}+V(q)+\frac{K_1(p_{\zeta})}{m_{\zeta}}
+\frac{K_2(p_{\xi})}{m_{\xi}}+k_BT(\zeta+\xi)
\\
&=H(q,p)+\frac{K_1(p_{\zeta})}{m_{\zeta}}
+\frac{K_2(p_{\xi})}{m_{\xi}}+k_BT(\zeta+\xi)\;,
\end{align}
\end{subequations}
where $(q,p)$ are the physical degrees of freedom (coordinates and momenta),
with mass $m$,
to be simulated at constant temperature $T$,
while $\zeta$ and $\xi$ are the BK `demons', with corresponding
inertial parameters $m_{\zeta}$ and $m_{\xi}$,
and associated momenta $(p_{\zeta},p_{\xi})$ \cite{Bulgac90,Kusnezov90,Kusnezov92}.
$K_1$ and $K_2$ provide the kinetic energy of demon variables,
and for the moment are left arbitrary.

Upon defining the phase space point as
$x=(q,\zeta,\xi,p,p_{\zeta},p_{\xi})=(x_1,x_2,x_3,x_4,x_5,x_6)$,
one can introduce an
antisymmetric BK tensor field as
\begin{equation}
\mbox{\boldmath $\cal  B$}^{\rm BK}
=
\left[
\begin{array}{cccccc}
0 & 0 & 0 & 1 & 0 & -G_2 \\
0 & 0 & 0 & 0 & \frac{\partial G_1}{\partial p} & 0 \\
0 & 0 & 0 & 0 & 0 & \frac{\partial G_2}{\partial q} \\
-1& 0 & 0 & 0 & -G_1 & 0 \\
0 & -\frac{\partial G_1}{\partial p}& 0 & G_1 & 0 & 0 \\
G_2 & 0 & -\frac{\partial G_2}{\partial q} & 0 & 0 & 0
\end{array}
\right]\;\label{eq:B-BK}
\end{equation}
where $G_1$ and $G_2$ are functions of system variables $(p, q)$ only.

Substituting $\mbox{\boldmath $\cal  B$}^{\rm BK}$ 
and $H^{\rm BK}$ into Eq.~(\ref{eq:gen-psf}),
we obtain the energy-conserving equations
\begin{subequations}
\label{eq:qdotBK}
\begin{align}
\dot{q}&=\frac{\partial H}{\partial p}-\frac{G_2(q,p)}{m_{\xi}}
\frac{\partial K_2}{\partial p_{\xi}}
\\
\dot{\zeta}&=\frac{1}{m_{\zeta}}\frac{\partial G_1}{\partial p}
\frac{\partial K_1}{\partial p_{\zeta}}
\\
\dot{\xi}&=\frac{1}{m_{\xi}}\frac{\partial G_2}{\partial q}
\frac{\partial K_2}{\partial p_{\xi}}
\\
\dot{p}&=-\frac{\partial H}{\partial q}-\frac{G_1(q,p)}{m_{\zeta}}
\frac{\partial K_1}{\partial p_{\zeta}}
\\
\dot{p}_{\zeta}&= G_1\frac{\partial H}{\partial p}
-k_BT\frac{\partial G_1}{\partial p}
\\
\dot{p}_{\xi}&=G_2\frac{\partial H}{\partial q}
-k_BT\frac{\partial G_2}{\partial q}\label{eq:pxidotBK}\;.
\end{align}
\end{subequations}
The associated invariant measure for the BK flow is discussed in Appendix \ref{app:BK-stat}.

\subsection{Algorithm for BK Dynamics}

In order to derive a measure preserving algorithms, the first step,
following Eq.~(\ref{eq:Hdecomp}), is to introduce
a splitting of $H^{\rm BK}$:
\begin{subequations}
\begin{align}
H_1^{\rm BK}&= V(q) \\
H_2^{\rm BK}&= \frac{p^2}{2m} \\
H_3^{\rm BK}&= k_BT\zeta \\
H_4^{\rm BK}&= k_BT\xi \\
H_5^{\rm BK}&= \frac{K_1(p_{\zeta})}{m_{\zeta}} \\
H_6^{\rm BK}&= \frac{K_2(p_{\xi})}{m_{\xi}}\;.
\end{align}
\end{subequations}
A measure-preserving splitting of the Liouville operator then follows
from Eq.~(\ref{eq:Ldecomp}):
\begin{subequations}
\begin{align}
L_1^{\rm BK}
& = -\frac{\partial V}{\partial q}\frac{\partial}{\partial p}
+G_2\frac{\partial V}{\partial q}\frac{\partial}{\partial p_{\xi}}
\\
L_2^{\rm BK}
&= \frac{p}{m}\frac{\partial}{\partial q}
+G_1\frac{p}{m}\frac{\partial}{\partial p_{\zeta}}
\\
L_3^{\rm BK}
&= -k_BT\frac{\partial G_1}{\partial p}\frac{\partial}{\partial p_{\zeta}}
\\
L_4^{\rm BK}
&= -k_BT\frac{\partial G_2}{\partial q}\frac{\partial}{\partial p_{\xi}}
\\
L_5^{\rm BK}
&=
\frac{1}{m_{\zeta}}\frac{\partial G_1}{\partial p}
\frac{\partial K_1}{\partial p_{\zeta}}\frac{\partial}{\partial \zeta}
-\frac{G_1}{m_{\zeta}}
\frac{\partial K_1}{\partial p_{\zeta}}\frac{\partial}{\partial p}
\\
L_6^{\rm BK}
& =
-\frac{G_2}{m_{\xi}}\frac{\partial K_2}{\partial p_{\xi}}
\frac{\partial}{\partial q}
+
\frac{1}{m_{\xi}}\frac{\partial G_2}{\partial q}
\frac{\partial K_2}{\partial p_{\xi}}
\frac{\partial}{\partial \xi}\;.
\end{align}
\end{subequations}
Upon choosing a symmetric Trotter factorization of
the BL Liouville operator based on the decomposition
\begin{equation}
L^{\rm BK}=\sum_{\alpha=1}^8L_{\alpha}^{\rm BK}
\end{equation}
a measure-preserving algorithm can be produced in full generality.

In practice, a choice of $K_1$, $K_2$, $G_1$, $G_2$ must be made in order
obtain explicit formulas.
In this paper, we make the following simple choices:
\begin{subequations}
\label{eq:choiceG1}
\begin{align}
G_1&= p \label{eq:choicea}\\
G_2&= q\\
K_1&= \frac{p_{\zeta}^2}{2}\\
K_2&= \frac{p_{\xi}^2}{2}\;.\label{eq:choiceb}
\end{align}
\end{subequations}
In terms of Eqs~(\ref{eq:choicea}--\ref{eq:choiceb}),
the antisymmetric BK tensor becomes
\begin{equation}
\tilde{\mbox{\boldmath $\cal  B$}}^{\rm BK}
=
\left[
\begin{array}{cccccc}
0 & 0 & 0 & 1 & 0 & -q \\
0 & 0 & 0 & 0 & 1 & 0  \\
0 & 0 & 0 & 0 & 0 & 1  \\
-1& 0 & 0 & 0 & -p & 0  \\
0 & -1& 0 & p & 0 & 0 \\
q & 0 & -1 & 0 & 0 & 0
\end{array}
\right]\;,
\end{equation}
and the Hamiltonian reads
\begin{eqnarray}
\tilde{H}^{\rm BK}
&=&H(q,p)+\frac{p_{\zeta}^2}{2m_{\zeta}}
+\frac{p_{\xi}^2}{2m_{\xi}}
+k_BT(\zeta+\xi)\;.
\end{eqnarray}
The split Liouville operators now simplify as follows:
\begin{subequations}
\label{eq:bcL1}
\begin{align}
\tilde{L}_1^{\rm BK}
&=-\frac{\partial V}{\partial q}\frac{\partial}{\partial p}
+q\frac{\partial V}{\partial q}\frac{\partial}{\partial p_{\xi}}
\\
\tilde{L}_2^{\rm BK}
&=\frac{p}{m}\frac{\partial}{\partial q}
+\frac{p^2}{m}\frac{\partial}{\partial p_{\zeta}}
\\
\tilde{L}_3^{\rm BK}
&=-k_BT\frac{\partial}{\partial p_{\zeta}}
\\
\tilde{L}_4^{\rm BK}
&=-k_BT\frac{\partial}{\partial p_{\xi}}
\\
\tilde{L}_5^{\rm BK}
&=
\frac{p_{\zeta}}{m_{\zeta}}\frac{\partial}{\partial \zeta}
-\frac{p_{\zeta}}{m_{\zeta}} p\frac{\partial}{\partial p}
+\frac{p_{\zeta}^2}{m_{\zeta}}\frac{\partial}{\partial p_{\eta}}
\\
\tilde{L}_6^{\rm BK}
&= -\frac{p_{\xi}}{m_{\xi}} q\frac{\partial}{\partial q}
+ \frac{p_{\xi}}{m_{\xi}} \frac{\partial}{\partial \xi}
+
\frac{p_{\xi}^2}{m_{\xi}} \frac{\partial}{\partial p_{\chi}}
\end{align}
\end{subequations}

For the purposes of defining an efficient integration algorithm,
we combine commuting Liouville operators as follows:
\begin{subequations}
\begin{align}
L_A^{\rm BK}&\equiv \tilde{L}_1^{\rm BK}+\tilde{L}_4^{\rm BK} \nonumber \\
  & = F(q)\frac{\partial}{\partial p}
+F_{p_{\xi}}\frac{\partial}{\partial p_{\xi}}
\\
L_B^{\rm BK}&\equiv \tilde{L}_2^{\rm BK}+\tilde{L}_3^{\rm BK} \nonumber\\
  &= \frac{p}{m}\frac{\partial}{\partial q}
+F_{p_{\zeta}}\frac{\partial}{\partial p_{\zeta}}
\\
L_C^{\rm BK}&\equiv \tilde{L}_5^{\rm BK}+\tilde{L}_6^{\rm BK}
\nonumber\\
&= -\frac{p_{\zeta}}{m_{\zeta}}p\frac{\partial}{\partial p}
-\frac{p_{\xi}}{m_{\xi}}q\frac{\partial}{\partial q}
+\frac{p_{\zeta}}{m_{\zeta}}\frac{\partial}{\partial \zeta}
+\frac{p_{\xi}}{m_{\xi}}\frac{\partial}{\partial \xi}
\end{align}
\end{subequations}
where
\begin{subequations}
\begin{align}
F(q)&= -\partial V/\partial q \\
F_{p_{\xi}}&= q\frac{\partial V}{\partial q}-k_BT
\\
F_{p_{\zeta}}&=\frac{p^2}{m}-k_BT \; .
\end{align}
\end{subequations}
Defining
\begin{equation}
U^{\rm BK}_{\alpha}(\tau)=\exp\left[\tau \tilde{L}^{\rm BK}_{\alpha}\right]\;,
\end{equation}
where $\alpha=A,B,C$,
one possible reversible measure-preserving integration
algorithm for the BK thermostat is then
\begin{eqnarray}
U(\tau)^{\rm BK}&=&
U_B^{\rm BK}\left(\frac{\tau}{4}\right)
U_C^{\rm BK}\left(\frac{\tau}{2}\right)
U_B^{\rm BK}\left(\frac{\tau}{4}\right)
\nonumber\\
&\times& U_A^{\rm BK}\left(\tau\right)
\nonumber\\
&\times&
U_B^{\rm BK}\left(\frac{\tau}{4}\right)
U_C^{\rm BK}\left(\frac{\tau}{2}\right)
U_B^{\rm BK}\left(\frac{\tau}{4}\right)
\;.
\end{eqnarray}

Using the so-called direct translation technique \cite{Tuckerman92} we can expand the above
symmetric break-up of the Liouville operator into a pseudo-code form,
ready to be implemented on the computer:
\begin{itemize}
\item
$
\left.
\begin{array}{ccl}
q&\to &q+\frac{\tau}{4} \frac{p}{m} \\
p_{\zeta} & \to &
p_{\zeta}+
\frac{\tau}{4}F_{p_{\zeta}}
\end{array}
\right\} :  U_B^{\rm BK}\left(\frac{\tau}{4}\right)
$
\item
$
\left.
\begin{array}{ccl}
p & \to & p\exp\left[-\frac{\tau}{2} \frac{p_{\zeta}}{m_{\zeta}}\right]\\
q &\to & q\exp\left[-\frac{\tau}{2}\frac{p_{\xi}}{m_{\xi}}\right]\\
\zeta&\to &\zeta+\frac{\tau}{2} \frac{p_{\zeta}}{m_{\zeta}} \\
\xi &\to & \xi +\frac{\tau}{2} \frac{p_{\xi}}{m_{\xi}}
\end{array}
\right\} :  U_C^{\rm BK}\left(\frac{\tau}{2}\right)
$
\item
$
\left.
\begin{array}{ccl}
q&\to &q+\frac{\tau}{4} \frac{p}{m} \\
p_{\zeta} & \to &
p_{\zeta}+
\frac{\tau}{4}F_{p_{\zeta}}
\end{array}
\right\} :  U_B^{\rm BK}\left(\frac{\tau}{4}\right)
$
\item $
\left.
\begin{array}{ccl}
p&\to &p+\tau F \\
p_{\xi} & \to &
p_{\xi}+
\tau F_{p_{\xi}}
\end{array}
\right\} : U_A^{\rm BK}(\tau)
$
\item
$
\left.
\begin{array}{ccl}
q&\to &q+\frac{\tau}{4} \frac{p}{m} \\
\eta &\to & \eta + \frac{\tau}{4} \frac{p_{\eta}}{m_{\eta}} \\
p_{\zeta} & \to &
p_{\zeta}+
\frac{\tau}{4}F_{p_{\zeta}}
\end{array}
\right\} :  U_B^{\rm BK}\left(\frac{\tau}{4}\right)
$
\item
$
\left.
\begin{array}{ccl}
p & \to & p\exp\left[-\frac{\tau}{2} \frac{p_{\zeta}}{m_{\zeta}}\right]\\
q &\to & q\exp\left[-\frac{\tau}{2}\frac{p_{\xi}}{m_{\xi}}\right]\\
\zeta&\to &\zeta+\frac{\tau}{2} \frac{p_{\zeta}}{m_{\zeta}} \\
\xi &\to & \xi +\frac{\tau}{2} \frac{p_{\xi}}{m_{\xi}}
\end{array}
\right\} :  U_C^{\rm BK}\left(\frac{\tau}{2}\right)
$
\item
$
\left.
\begin{array}{ccl}
q&\to &q+\frac{\tau}{4} \frac{p}{m} \\
p_{\zeta} & \to &
p_{\zeta}+
\frac{\tau}{4}F_{p_{\zeta}}
\end{array}
\right\} :  U_B^{\rm BK}\left(\frac{\tau}{4}\right)
$
\end{itemize}



\newpage

\section{Bulgac-Kusnezov-Nos\'e-Hoover dynamics}
\label{sec:cbk}

The  BKNH Hamiltonian
\begin{equation}
H^{\rm BKNH} = H(q,p)+\frac{K_1(p_{\zeta})}{m_{\zeta}}
+\frac{K_2(p_{\xi})}{m_{\xi}}+\frac{p_{\eta}^2}{2m_{\eta}}
+k_BT(\zeta+\xi) +2k_BT\eta
\;\label{eq:H-CBK}
\end{equation}
is simply the BK Hamiltonian augmented by the Nos\'e variables
$(\eta,p_{\eta})$ with mass $m_{\eta}$.
With the antisymmetric BKNH tensor
\begin{equation}
\mbox{\boldmath $\cal  B$}^{\rm BKNH}
=
\left[
\begin{array}{cccccccc}
0  & 0 & 0 &  0 & 1 & 0 & -G_2 & 0 \\
0  & 0 & 0 &  0 & 0 & \frac{\partial G_1}{\partial p} & 0 & 0  \\
0  & 0 & 0 &  0 & 0 & 0 & \frac{\partial G_2}{\partial q} & 0  \\
0  & 0 & 0 &  0 & 0 & 0 & 0 & 1  \\
-1  & 0 & 0 &  0 & 0 &  -G_1 & 0 & 0  \\
 0 & -\frac{\partial G_1}{\partial p}& 0 &  0 & G_1 & 0 & 0 & -p_{\zeta}  \\
G_2 & 0 & -\frac{\partial G_2}{\partial q} &  0 & 0 & 0 & 0 & -p_{\xi} \\
0 & 0 & 0 & -1 & 0 & p_{\zeta} & p_{\xi} & 0  \\
\end{array}
\right]\;.
\label{eq:B-CBK}
\end{equation}
we obtain from Eq.~(\ref{eq:gen-psf}) equations of motion
for the  phase space variables $x=(q,\zeta,\xi,\eta,p,p_{\zeta},p_{\xi},p_{\eta})=
(x_1,x_2,x_3,x_4,x_5,x_6,x_7,x_8)$:
\begin{subequations}
\begin{align}
\dot{q}&=\frac{\partial H}{\partial p}-\frac{G_2(q,p)}{m_{\xi}}
\frac{\partial K_2}{\partial p_{\xi}}
\\
\dot{\zeta}&=\frac{1}{m_{\zeta}}\frac{\partial G_1}{\partial p}
\frac{\partial K_1}{\partial p_{\zeta}}
\\
\dot{\xi}&=\frac{1}{m_{\xi}}\frac{\partial G_2}{\partial q}
\frac{\partial K_2}{\partial p_{\xi}}
\\
\dot{\eta}&=\frac{p_{\eta}}{m_{\eta}}
\\
\dot{p}&=-\frac{\partial H}{\partial q}-\frac{G_1(q,p)}{m_{\zeta}}
\frac{\partial K_1}{\partial p_{\zeta}}
\\
\dot{p}_{\zeta}&=G_1\frac{\partial H}{\partial p}
-k_BT\frac{\partial G_1}{\partial p}-p_{\zeta}\frac{p_{\eta}}{m_{\eta}}
\\
\dot{p}_{\xi}&=G_2\frac{\partial H}{\partial q}
-k_BT\frac{\partial G_2}{\partial q}-p_{\xi}\frac{p_{\eta}}{m_{\eta}}
\\
\dot{p}_{\eta}&=
\frac{p_{\zeta}}{m_{\zeta}}
\frac{\partial K_1}{\partial p_{\zeta}}
+\frac{p_{\xi}}{m_{\xi}}
\frac{\partial K_2}{\partial p_{\xi}}
-2k_BT
\;.
\end{align}
\end{subequations}
Here, a single Nos\'{e} variable is coupled to both of the BK demons $\zeta$ and $\xi$.
The associated invariant measure is discussed in Appendix \ref{app:CBK-stat}.


\subsection{Algorithm for BKNH dynamics}\label{sec:CBK-alg}

The Hamiltonian can be split as
\begin{subequations}
\begin{align}
H_1^{\rm BKNH}&= V(q) \\
H_2^{\rm BKNH}&= \frac{p^2}{2m} \\
H_3^{\rm BKNH}&= k_BT\zeta \\
H_4^{\rm BKNH}&= k_BT\xi \\
H_5^{\rm BKNH}&= \frac{K_1(p_{\zeta})}{m_{\zeta}} \\
H_6^{\rm BKNH}&= \frac{K_2(p_{\xi})}{m_{\xi}}\\
H_7^{\rm BKNH}&= \frac{p^2_{\eta}}{2m_{\eta}} \\
H_8^{\rm BKNH}&=  2k_BT\eta
\end{align}
\end{subequations}
The measure-preserving splitting~\cite{Ezra06} of the Liouville operator
\begin{equation}
L_{\alpha}={\cal B}_{ij}^{\rm BKNH}
\frac{\partial H^{\rm BKNH}_{\alpha}}{\partial x_j}
\frac{\partial}{\partial x_i}
\end{equation}
yields
\begin{subequations}
\begin{align}
L_1^{\rm BKNH}
&=-\frac{\partial V}{\partial q}\frac{\partial}{\partial p}
+G_2\frac{\partial V}{\partial q}\frac{\partial}{\partial p_{\xi}}
\\
L_2^{\rm BKNH}
&=\frac{p}{m}\frac{\partial}{\partial q}
+G_1\frac{p}{m}\frac{\partial}{\partial p_{\zeta}}
\\
L_3^{\rm BKNH}
&= -k_BT\frac{\partial G_1}{\partial p}\frac{\partial}{\partial p_{\zeta}}
\\
L_4^{\rm BKNH}
&= -k_BT\frac{\partial G_2}{\partial q}\frac{\partial}{\partial p_{\xi}}
\\
L_5^{\rm BKNH}
&=
\frac{1}{m_{\zeta}}\frac{\partial G_1}{\partial p}
\frac{\partial K_1}{\partial p_{\zeta}}\frac{\partial}{\partial \zeta}
-\frac{G_1}{m_{\zeta}}
\frac{\partial K_1}{\partial p_{\zeta}}\frac{\partial}{\partial p}
+\frac{p_{\zeta}}{m_{\zeta}}
\frac{\partial K_1}{\partial p_{\zeta}}\frac{\partial}{\partial p_{\eta}}
\\
L_6^{\rm BKNH}
&=
-\frac{G_2}{m_{\xi}}\frac{\partial K_2}{\partial p_{\xi}}
\frac{\partial}{\partial q}
+
\frac{1}{m_{\xi}}\frac{\partial G_2}{\partial q}
\frac{\partial K_2}{\partial p_{\xi}}
\frac{\partial}{\partial \xi}
+\frac{p_{\xi}}{m_{\xi}}\frac{\partial K_2}{\partial p_{\xi}}
\frac{\partial}{\partial p_{\eta}}
\\
L_7^{\rm BKNH}
&=
\frac{p_{\eta}}{m_{\eta}}\frac{\partial}{\partial \eta}
- \frac{p_{\eta}}{m_{\eta}}p_{\zeta}
\frac{\partial}{\partial p_{\zeta}}
- \frac{p_{\eta}}{m_{\eta}}p_{\xi}
\frac{\partial}{\partial p_{\xi}}
\\
L_8^{\rm BKNH}
&=-2k_BT\frac{\partial}{\partial p_{\eta}}
\;.
\end{align}
\end{subequations}

At this stage, we leave the general formulation and adopt
the particular choice of $K_1$, $K_2$, $G_1$, and $G_2$
given in Eq.~(\ref{eq:choiceG1}).
The antisymmetric BKNH tensor becomes
\begin{equation}
\tilde{\mbox{\boldmath $\cal  B$}}^{\rm BKNH}
=
\left[
\begin{array}{cccccccc}
0  & 0 & 0 &  0 & 1 & 0 & -q & 0 \\
0  & 0 & 0 &  0 & 0 & 1 & 0 & 0  \\
0  & 0 & 0 &  0 & 0 & 0 & 1 & 0  \\
0  & 0 & 0 &  0 & 0 & 0 & 0 & 1  \\
-1  & 0 & 0 &  0 & 0 &  -p & 0 & 0  \\
0 & -1& 0 &  0 & p & 0 & 0 & -p_{\zeta}  \\
q & 0 & -1 &  0 & 0 & 0 & 0 & -p_{\xi} \\
0 & 0 & 0 & -1 & 0 & p_{\zeta} & p_{\xi} & 0  \\
\end{array}
\right]\;,
\end{equation}
and the Hamiltonian simplifies to
\begin{equation}
\tilde{H}^{\rm BKNH}
= H(q,p)+\frac{p_{\zeta}^2}{2m_{\zeta}}
+\frac{p_{\xi}^2}{2m_{\xi}}+\frac{p_{\eta}^2}{2m_{\eta}}
+k_BT(\zeta+\xi)+2_BT\eta\;.
\end{equation}
The split Liouville operators are now
\begin{subequations}
\begin{align}
\tilde{L}_1^{\rm BKNH}
&=-\frac{\partial V}{\partial q}\frac{\partial}{\partial p}
+q\frac{\partial V}{\partial q}\frac{\partial}{\partial p_{\xi}}
\label{eq:cL1}
\\
\tilde{L}_2^{\rm BKNH}
&=\frac{p}{m}\frac{\partial}{\partial q}
+\frac{p^2}{m}\frac{\partial}{\partial p_{\zeta}}
\\
\tilde{L}_3^{\rm BKNH}
&=-k_BT\frac{\partial}{\partial p_{\zeta}}
\\
\tilde{L}_4^{\rm BKNH}
&=-k_BT\frac{\partial}{\partial p_{\xi}}
\\
\tilde{L}_5^{\rm BKNH}
&=
\frac{p_{\zeta}}{m_{\zeta}} \frac{\partial}{\partial \zeta}
-\frac{p_{\zeta}}{m_{\zeta}} p\frac{\partial}{\partial p}
+\frac{p_{\zeta}^2}{m_{\zeta}}
\frac{\partial}{\partial p_{\eta}}
\\
\tilde{L}_6^{\rm BKNH}
&=
-\frac{p_{\xi}}{m_{\xi}}
q\frac{\partial}{\partial q}
+
\frac{p_{\xi}}{m_{\xi}}
\frac{\partial}{\partial \xi}
+\frac{p_{\xi}^2}{m_{\xi}}
\frac{\partial}{\partial p_{\eta}}
\\
\tilde{L}_7^{\rm BKNH}
&=
\frac{p_{\eta}}{m_{\eta}}\frac{\partial}{\partial \eta}
- \frac{p_{\eta}}{m_{\eta}}p_{\zeta}
\frac{\partial}{\partial p_{\zeta}}
- \frac{p_{\eta}}{m_{\eta}}p_{\xi}
\frac{\partial}{\partial p_{\xi}}
\\
\tilde{L}_8^{\rm BKNH}
&= -2k_BT\frac{\partial}{\partial p_{\eta}}
\;.  \label{eq:cL10}
\end{align}
\end{subequations}
For the purposes of defining an efficient integration algorithm,
we combine commuting Liouville operators as follows:
\begin{subequations}
\begin{align}
L_A^{\rm BKNH}&\equiv \tilde{L}_1^{\rm BKNH}+\tilde{L}_4^{\rm BKNH}+\tilde{L}_7^{\rm BKNH} \nonumber \\
  & = F(q)\frac{\partial}{\partial p}
           +\frac{p_{\eta}}{m_{\eta}}\frac{\partial}{\partial\eta}
-\frac{p_{\eta}}{m_{\eta}}p_{\zeta}\frac{\partial}{\partial p_{\zeta}}
+\left(-\frac{p_{\chi}}{m_{\chi}}p_{\xi}+F_{p_{\xi}}\right)\frac{\partial}{\partial p_{\xi}}
\\
\tilde{L}_B^{\rm BKNH}&\equiv \tilde{L}_2^{\rm BKNH}+\tilde{L}_3^{\rm BKNH} \nonumber\\
  &= \frac{p}{m}\frac{\partial}{\partial q}
+F_{p_{\zeta}}\frac{\partial}{\partial p_{\zeta}}
\\
L_C^{\rm BKNH}&\equiv \tilde{L}_5^{\rm BKNH}+\tilde{L}_6^{\rm BKNH} +\tilde{L}_8^{\rm BKNH}
\nonumber\\
&= -\frac{p_{\zeta}}{m_{\zeta}}p\frac{\partial}{\partial p}
-\frac{p_{\xi}}{m_{\xi}}q\frac{\partial}{\partial q}
+\frac{p_{\zeta}}{m_{\zeta}}\frac{\partial}{\partial \zeta}
+\frac{p_{\xi}}{m_{\xi}}\frac{\partial}{\partial \xi}
+F_{p_{\eta}}\frac{\partial}{p_{\eta}}
\;,
\end{align}
\end{subequations}
where
\begin{subequations}
\begin{align}
F(q)&= -\frac{\partial V}{\partial q} \\
F_{p_{\xi}}&=q\frac{\partial V}{\partial q}-k_BT
\\
F_{p_{\zeta}}&=\frac{p^2}{m}-k_BT
\\
F_{p_{\eta}}&=\frac{p_{\zeta}^2}{m_{\zeta}}+\frac{p_{\xi}^2}{m_{\xi}}
-2k_BT
\;.
\end{align}
\end{subequations}

In $L_A$ there appears an operator with the form
\begin{equation}
\label{eq:operator_CBK}
L_i=\left(-\frac{p_k}{m_k}p_i+F_{p_i}\right)\frac{\partial}{\partial p_i}
\;,
\end{equation}
where $(k,i)=(\chi,\xi)$ for $L_A$.
The action of the propagator associated with this operator on $p_i$
is derived in Appendix~\ref{app:app1}, and is given by
\begin{equation}
e^{\tau L_i}p_i
=
p_ie^{-\tau \frac{p_k}{m_k}}+
\tau F_{p_i}e^{-\tau \frac{p_k}{2m_k}}
\left(\tau \frac{p_k}{2m_k}\right)^{-1}
\sinh\left[\tau \frac{p_k}{2m_k}\right]
\;.
\end{equation}
The apparently singular function
\begin{equation}
\left(\tau\frac{p_k}{2m_k}\right)^{-1}
\sinh\left[\tau\frac{p_k}{2m_k}\right]
\end{equation}
is in fact well behaved as $p_k \to 0$, and
can be expanded in a Maclaurin series to suitably high order~\cite{Martyna96}.
In our implementation we used an eighth order expansion.

The propagators for the BKNH dynamics can now be defined as
\begin{equation}
U_{\alpha}^{\rm BKNH}(\tau)=\exp\left[\tau \tilde{L}_{\alpha}^{\rm BKNH}\right]
\;,
\end{equation}
where $\alpha=A,B,C$.
One possible reversible measure-preserving integration
algorithm for the BKNH thermostat can then be derived from
the following Trotter factorization:
\begin{eqnarray}
U(\tau)^{\rm BKNH}&=&
U_B^{\rm BKNH}\left(\frac{\tau}{4}\right)
U_C^{\rm BKNH}\left(\frac{\tau}{2}\right)
U_B^{\rm BKNH}\left(\frac{\tau}{4}\right)
\nonumber\\
&\times& U_A^{\rm BKNH}\left(\tau\right)
\nonumber\\
&\times&
U_B^{\rm BKNH}\left(\frac{\tau}{4}\right)
U_C^{\rm BKNH}\left(\frac{\tau}{2}\right)
U_B^{\rm BKNH}\left(\frac{\tau}{4}\right)
\;.
\end{eqnarray}
The direct translation technique gives the following pseudo-code:
\begin{itemize}
\item
$
\left.
\begin{array}{ccl}
q&\to &q+\frac{\tau}{4} \frac{p}{m} \\
p_{\zeta} & \to &
p_{\zeta}+ \frac{\tau}{4}F_{p_{\zeta}}
\end{array}
\right\} :  U_B^{\rm BKNH}\left(\frac{\tau}{4}\right)
$
\item
$
\left.
\begin{array}{ccl}
p & \to & p\exp\left[-\frac{\tau}{2} \frac{p_{\zeta}}{m_{\zeta}}\right]\\
q &\to & q\exp\left[-\frac{\tau}{2}\frac{p_{\xi}}{m_{\xi}}\right]\\
\zeta&\to &\zeta+\frac{\tau}{2} \frac{p_{\zeta}}{m_{\zeta}} \\
\xi &\to & \xi +\frac{\tau}{2} \frac{p_{\xi}}{m_{\xi}} \\
p_{\eta} & \to & p_{\eta} +\frac{\tau}{2}F_{p_{\zeta}} \\
\end{array}
\right\} :  U_C^{\rm BKNH}\left(\frac{\tau}{2}\right)
$
\item
$
\left.
\begin{array}{ccl}
q&\to &q+\frac{\tau}{4} \frac{p}{m} \\
p_{\zeta} & \to &
p_{\zeta}+ \frac{\tau}{4}F_{p_{\zeta}}
\end{array}
\right\} :  U_B^{\rm BKNH}\left(\frac{\tau}{4}\right)
$
\item $
\left.
\begin{array}{ccl}
p&\to &p+\tau F(q) \\
p_{\xi} & \to &
p_{\xi}
+
\tau F_{p_{\xi}}
\\
\eta &\to & \eta + \tau\frac{p_{\eta}}{m_{\eta}}
\\
p_{\zeta} &\to& p_{\zeta}\exp\left[-\tau\frac{p_{\eta}}{m_{\eta}}\right]
\end{array}
\right\} : U_A^{\rm BKNH}(\tau)
$
\item
$
\left.
\begin{array}{ccl}
q&\to &q+\frac{\tau}{4} \frac{p}{m} \\
p_{\zeta} & \to &
p_{\zeta}+
\frac{\tau}{4}F_{p_{\zeta}}
\end{array}
\right\} :  U_B^{\rm BKNH}\left(\frac{\tau}{4}\right)
$
\item
$
\left.
\begin{array}{ccl}
p & \to & p\exp\left[-\frac{\tau}{2} \frac{p_{\zeta}}{m_{\zeta}}\right]\\
q &\to & q\exp\left[-\frac{\tau}{2}\frac{p_{\xi}}{m_{\xi}}\right]\\
\zeta&\to &\zeta+\frac{\tau}{2} \frac{p_{\zeta}}{m_{\zeta}} \\
\xi &\to & \xi +\frac{\tau}{2} \frac{p_{\xi}}{m_{\xi}} \\
p_{\eta} & \to & p_{\eta} +\frac{\tau}{2}F_{p_{\eta}} \\
\end{array}
\right\} :  U_C^{\rm BKNH}\left(\frac{\tau}{2}\right)
$
\item
$
\left.
\begin{array}{ccl}
q&\to &q+\frac{\tau}{4} \frac{p}{m} \\
p_{\zeta} & \to &
p_{\zeta}+ \frac{\tau}{4}F_{p_{\zeta}}
\end{array}
\right\} :  U_B^{\rm BKNH}\left(\frac{\tau}{4}\right)
$
\end{itemize}


\newpage

\section{Bulgac-Kusnezov-Nos\'e-Hoover chain}
\label{sec:nbkc}

For simplicity, we explicitly treat only
the case in which the  $p_{\zeta}$ and $p_{\xi}$
demons are each coupled to a standard NH thermostat (length one).
It would be straightforward to couple each of the demons to NH chains \cite{Martyna92},
and the general case can be easily inferred from what follows.
Define the Hamiltonian
\begin{eqnarray}
H^{\rm BKNHC}
&=&H(q,p)+\frac{K_1(p_{\zeta})}{m_{\zeta}}
+\frac{K_2(p_{\xi})}{m_{\xi}}+\frac{p_{\eta}^2}{2m_{\eta}}
+\frac{p_{\chi}^2}{2m_{\chi}}
+k_BT(\zeta+\xi+\eta+\chi)\;.
\label{eq:H-NBK}
\end{eqnarray}
Upon defining the phase space point $x=(q,\zeta,\xi,\eta,\chi,p,p_{\zeta},p_{\xi},p_{\eta},p_{\chi})
=(x_1,x_2,x_3,x_4,x_5,x_6,x_7,x_8,x_9,x_{10})$
and the antisymmetric BKNHC tensor
\begin{equation}
\mbox{\boldmath $\cal  B$}^{\rm BKNHC}
=
\left[
\begin{array}{cccccccccc}
0 & 0 & 0 & 0 & 0 & 1 & 0 & -G_2 & 0 & 0\\
0 & 0 & 0 & 0 & 0 & 0 & \frac{\partial G_1}{\partial p} & 0 & 0 & 0 \\
0 & 0 & 0 & 0 & 0 & 0 & 0 & \frac{\partial G_2}{\partial q} & 0 & 0 \\
0 & 0 & 0 & 0 & 0 & 0 & 0 & 0 & 1 & 0 \\
0 & 0 & 0 & 0 & 0 & 0 & 0 & 0 & 0 & 1 \\
-1& 0 & 0 & 0 & 0 & 0 &  -G_1 & 0 & 0 & 0 \\
0 & -\frac{\partial G_1}{\partial p}& 0 & 0 & 0 & G_1 & 0 & 0 & -p_{\zeta} & 0 \\
G_2 & 0 & -\frac{\partial G_2}{\partial q} & 0 & 0 & 0 & 0 & 0 & 0 & -p_{\xi}\\
0 & 0 & 0 & -1 & 0 & 0 & p_{\zeta} & 0 & 0 & 0\\
0 & 0 & 0 & 0 & -1 & 0 & 0 & p_{\xi} & 0 & 0\\
\end{array}
\right]\;,
\label{eq:B-NBK}
\end{equation}
associated non-Hamiltonian equations of motion 
are 
\begin{equation}
\dot{x}_i =  {\cal  B}^{\rm BKNHC}_{ij} \, \frac{\partial H^{{\rm BKNHC}}}{\partial x_j}
\end{equation}
with $i =1, \ldots, 10$.

\subsection{Algorithm for BKNHC chain dynamics}

Splitting the BKNHC chain Hamiltonian as
\begin{subequations}
\begin{align}
H_1^{\rm BKNHC}&= V(q) \\
H_2^{\rm BKNHC}&= \frac{p^2}{2m} \\
H_3^{\rm BKNHC}&= k_BT\zeta \\
H_4^{\rm BKNHC}&= k_BT\xi \\
H_5^{\rm BKNHC}&= \frac{K_1(p_{\zeta})}{m_{\zeta}} \\
H_6^{\rm BKNHC}&= \frac{K_2(p_{\xi})}{m_{\xi}}\\
H_7^{\rm BKNHC}&= \frac{p^2_{\eta}}{2m_{\eta}} \\
H_8^{\rm BKNHC}&= k_BT\eta\\
H_9^{\rm BKNHC}&= \frac{p^2_{\chi}}{2m_{\chi}} \\
H_{10}^{\rm BKNHC}&= k_BT\chi
\;,
\end{align}
\end{subequations}
we obtain the corresponding 
measure-preserving splitting of the Liouville operator
\begin{equation}
L_{\alpha}={\cal B}_{ij}^{\rm BKNHC}
\frac{\partial H^{\rm BKNHC}_{\alpha}}{\partial x_j}
\frac{\partial}{\partial x_i} .
\end{equation}

At this stage we go directly to Eqs~(\ref{eq:choiceG1}).
The antisymmetric Nos\'e-Hoover-Bulgac-Kusnezov tensor becomes
\begin{equation}
\tilde{\mbox{\boldmath $\cal  B$}}^{\rm BKNHC}
=
\left[
\begin{array}{cccccccccc}
0 & 0 & 0 & 0 & 0 & 1 & 0 & -q & 0 & 0\\
0 & 0 & 0 & 0 & 0 & 0 & 1 & 0 & 0 & 0 \\
0 & 0 & 0 & 0 & 0 & 0 & 0 & 1 & 0 & 0 \\
0 & 0 & 0 & 0 & 0 & 0 & 0 & 0 & 1 & 0 \\
0 & 0 & 0 & 0 & 0 & 0 & 0 & 0 & 0 & 1 \\
-1& 0 & 0 & 0 & 0 & 0 &  -p & 0 & 0 & 0 \\
0 & -1& 0 & 0 & 0 & p & 0 & 0 & -p_{\zeta} & 0 \\
q & 0 & -1 & 0 & 0 & 0 & 0 & 0 & 0 & -p_{\xi}\\
0 & 0 & 0 & -1 & 0 & 0 & p_{\zeta} & 0 & 0 & 0\\
0 & 0 & 0 & 0 & -1 & 0 & 0 & p_{\xi} & 0 & 0\\
\end{array}
\right]\; ,
\end{equation}
the Hamiltonian
\begin{eqnarray}
\tilde{H}^{\rm BKNHC}
&=&H(q,p)+\frac{p_{\zeta}^2}{2m_{\zeta}}
+\frac{p_{\xi}^2}{2m_{\xi}}+\frac{p_{\eta}^2}{2m_{\eta}}
+\frac{p_{\chi}^2}{2m_{\chi}}
+k_BT(\zeta+\xi+\eta+\chi)\; 
\end{eqnarray}
and associated Liouville operators
\begin{subequations}
\label{eq:L}
\begin{align}
\tilde{L}_1^{\rm BKNHC}
&=-\frac{\partial V}{\partial q}\frac{\partial}{\partial p}
+q\frac{\partial V}{\partial q}\frac{\partial}{\partial p_{\xi}}
\label{eq:L1}
\\
\tilde{L}_2^{\rm BKNHC}
&=\frac{p}{m}\frac{\partial}{\partial q}
+\frac{p^2}{m}\frac{\partial}{\partial p_{\zeta}}
\\
\tilde{L}_3^{\rm BKNHC}
&= -k_BT\frac{\partial}{\partial p_{\zeta}}
\\
\tilde{L}_4^{\rm BKNHC}
&= -k_BT\frac{\partial}{\partial p_{\xi}}
\\
\tilde{L}_5^{\rm BKNHC}
&=
\frac{p_{\zeta}}{m_{\zeta}}\frac{\partial}{\partial \zeta}
-\frac{p_{\zeta}}{m_{\zeta}} p\frac{\partial}{\partial p}
+\frac{p_{\zeta}^2}{m_{\zeta}}\frac{\partial}{\partial p_{\eta}}
\\
\tilde{L}_6^{\rm BKNHC}
&= -\frac{p_{\xi}}{m_{\xi}} q\frac{\partial}{\partial q}
+ \frac{p_{\xi}}{m_{\xi}} \frac{\partial}{\partial \xi}
+
\frac{p_{\xi}^2}{m_{\xi}} \frac{\partial}{\partial p_{\chi}}
\\
\tilde{L}_7^{\rm BKNHC}
&=
\frac{p_{\eta}}{m_{\eta}}\frac{\partial}{\partial \eta}
- \frac{p_{\eta}}{m_{\eta}}p_{\zeta}
\frac{\partial}{\partial p_{\zeta}}
\\
\tilde{L}_8^{\rm BKNHC}
&= -k_BT\frac{\partial}{\partial p_{\eta}}
\\
\tilde{L}_{9}^{\rm BKNHC}
&=
\frac{p_{\chi}}{m_{\chi}}\frac{\partial}{\partial \chi}
-\frac{p_{\chi}}{m_{\chi}}p_{\xi}
\frac{\partial}{\partial p_{\xi}}
\\
\tilde{L}_{10}^{\rm BKNHC}
&=-k_BT\frac{\partial}{\partial p_{\chi}}
\;.
\label{eq:L10}
\end{align}
\end{subequations}

We combine commuting Liouville operators as follows:
\begin{subequations}
\begin{align}
L_A^{\rm BKNHC}&\equiv \tilde{L}_1^{\rm BKNHC}+\tilde{L}_4^{\rm BKNHC}
+\tilde{L}_9^{\rm BKNHC} \nonumber \\
  & = F(q) \frac{\partial}{\partial p}
           +\frac{p_{\chi}}{m_{\chi}}\frac{\partial}{\partial\chi}
+\left(-\frac{p_{\chi}}{m_{\chi}}p_{\xi}+F_{p_{\xi}}\right)\frac{\partial}{\partial p_{\xi}}
\\
L_B^{\rm BKNHC}&\equiv \tilde{L}_2^{\rm BKNHC}
+\tilde{L}_3^{\rm BKNHC} +\tilde{L}_7^{\rm BKNHC} \nonumber\\
  &= \frac{p}{m}\frac{\partial}{\partial q}
+\frac{p_{\eta}}{m_{\eta}}\frac{\partial}{\partial\eta}
+\left(-\frac{p_{\eta}}{m_{\eta}}p_{\zeta}+F_{p_{\zeta}}\right)\frac{\partial}{\partial p_{\zeta}}
\\
L_C^{\rm BKNHC}&\equiv \tilde{L}_5^{\rm BKNHC}+\tilde{L}_6^{\rm BKNHC}
+\tilde{L}_8^{\rm BKNHC}+\tilde{L}_{10}^{\rm BKNHC}
\nonumber\\
&= -\frac{p_{\zeta}}{m_{\zeta}}p\frac{\partial}{\partial p}
-\frac{p_{\xi}}{m_{\xi}}q\frac{\partial}{\partial q}
+\frac{p_{\zeta}}{m_{\zeta}}\frac{\partial}{\partial \zeta}
+\frac{p_{\xi}}{m_{\xi}}\frac{\partial}{\partial \xi}
+F_{p_{\eta}}\frac{\partial}{p_{\eta}}
+F_{p_{\chi}}\frac{\partial}{p_{\chi}}
\;,
\end{align}
\end{subequations}
where
\begin{subequations}
\begin{align}
F(q)&=-\frac{\partial V}{\partial q} \\
F_{p_{\xi}}&=q\frac{\partial V}{\partial q}-k_BT
\\
F_{p_{\zeta}}&=\frac{p^2}{m}-k_BT
\\
F_{p_{\eta}}&=\frac{p_{\zeta}^2}{m_{\zeta}}-k_BT
\\
F_{p_{\chi}}&=\frac{p_{\xi}^2}{m_{\xi}}-k_BT
\;.
\end{align}
\end{subequations}
Both in $L_A^{\rm BKNHC}$ and $L_B^{\rm BKNHC}$ there appears an operator with the form
\begin{equation}
L_i=\left(-\frac{p_k}{m_k}p_i+F_{p_i}\right)\frac{\partial}{\partial p_i}
\;,
\end{equation}
where $(k,i)=(\chi,\xi)$ for $L_A$ and
$(k,i)=(\eta,\zeta)$ for $L_B$.
Again following the derivation in Appendix~\ref{app:app1}, we find
\begin{equation}
e^{\tau L_i}p_i
=
p_ie^{-\tau \frac{p_k}{m_k}}+
\tau F_{p_i}e^{-\tau \frac{p_k}{2m_k}}
\left(\tau \frac{p_k}{2m_k}\right)^{-1}
\sinh\left[\tau \frac{p_k}{2m_k}\right]
\;.
\end{equation}
The function
$\left(\tau\frac{p_k}{2m_k}\right)^{-1}
\sinh\left[\tau\frac{p_k}{2m_k}\right]$
is treated through an eighth order expansion~\cite{Martyna96}.


The propagators
\begin{equation}
U_{\alpha}^{\rm BKNHC}(\tau)=\exp\left[\tau \tilde{L}_{\alpha}^{\rm BKNHC}\right]
\end{equation}
with $\alpha=A,B,C$ can now be introduced.
One possible reversible measure-preserving integration
algorithm for the BKNHC chain thermostat is then
\begin{eqnarray}
U(\tau)^{\rm BKNHC}&=&
U_B^{\rm BKNHC}\left(\frac{\tau}{4}\right)
U_C^{\rm BKNHC}\left(\frac{\tau}{2}\right)
U_B^{\rm BKNHC}\left(\frac{\tau}{4}\right)
\nonumber\\
&\times& U_A^{\rm BKNHC}\left(\tau\right)
\nonumber\\
&\times&
U_B^{\rm BKNHC}\left(\frac{\tau}{4}\right)
U_C^{\rm BKNHC}\left(\frac{\tau}{2}\right)
U_B^{\rm BKNHC}\left(\frac{\tau}{4}\right)
\;.
\end{eqnarray}
In pseudo-code form, we have the resulting integration algorithm:
\begin{itemize}
\item
$
\left.
\begin{array}{ccl}
q&\to &q+\frac{\tau}{4} \frac{p}{m} \\
\eta &\to & \eta + \frac{\tau}{4} \frac{p_{\eta}}{m_{\eta}} \\
p_{\zeta} & \to &
p_{\zeta}e^{-\frac{\tau}{4} \frac{p_{\eta}}{m_{\eta}}}+
\frac{\tau}{4}F_{p_{\zeta}}e^{-\frac{\tau}{4} \frac{p_{\eta}}{2m_{\eta}}}
\left(\frac{\tau}{4}\frac{p_{\eta}}{2m_{\eta}}\right)^{-1}
\sinh\left[\frac{\tau}{4}\frac{p_{\eta}}{2m_{\eta}}\right]
\end{array}
\right\} :  U_B^{\rm BKNHC}\left(\frac{\tau}{4}\right)
$
\item
$
\left.
\begin{array}{ccl}
p & \to & p\exp\left[-\frac{\tau}{2} \frac{p_{\zeta}}{m_{\zeta}}\right]\\
q &\to & q\exp\left[-\frac{\tau}{2}\frac{p_{\xi}}{m_{\xi}}\right]\\
\zeta&\to &\zeta+\frac{\tau}{2} \frac{p_{\zeta}}{m_{\zeta}} \\
\xi &\to & \xi +\frac{\tau}{2} \frac{p_{\xi}}{m_{\xi}} \\
p_{\eta} & \to & p_{\eta} +\frac{\tau}{2}F_{p_{\zeta}} \\
p_{\chi} &\to & p_{\chi} +\frac{\tau}{2} F_{p_{\chi}}
\end{array}
\right\} :  U_C^{\rm BKNHC}\left(\frac{\tau}{2}\right)
$
\item
$
\left.
\begin{array}{ccl}
q&\to &q+\frac{\tau}{4} \frac{p}{m} \\
\eta &\to & \eta + \frac{\tau}{4} \frac{p_{\eta}}{m_{\eta}} \\
p_{\zeta} & \to &
p_{\zeta}e^{-\frac{\tau}{4} \frac{p_{\eta}}{m_{\eta}}}+
\frac{\tau}{4}F_{p_{\zeta}}e^{-\frac{\tau}{4} \frac{p_{\eta}}{2m_{\eta}}}
\left(\frac{\tau}{4}\frac{p_{\eta}}{2m_{\eta}}\right)^{-1}
\sinh\left[\frac{\tau}{4}\frac{p_{\eta}}{2m_{\eta}}\right]
\end{array}
\right\} :  U_B^{\rm BKNHC}\left(\frac{\tau}{4}\right)
$
\item $
\left.
\begin{array}{ccl}
p&\to &p+\tau F \\
\chi &\to & \chi +\tau \frac{p_{\chi}}{m_{\chi}} \\
p_{\xi} & \to &
p_{\xi}e^{-\tau \frac{p_{\chi}}{m_{\chi}}}+
\tau F_{p_{\xi}}e^{-\tau \frac{p_{\chi}}{2m_{\chi}}}
\left(\tau\frac{p_{\chi}}{2m_{\chi}}\right)^{-1}
\sinh\left[\tau\frac{p_{\chi}}{2m_{\chi}}\right]
\end{array}
\right\} : U_A^{\rm BKNHC}(\tau)
$
\item
$
\left.
\begin{array}{ccl}
q&\to &q+\frac{\tau}{4} \frac{p}{m} \\
\eta &\to & \eta + \frac{\tau}{4} \frac{p_{\eta}}{m_{\eta}} \\
p_{\zeta} & \to &
p_{\zeta}e^{-\frac{\tau}{4} \frac{p_{\eta}}{m_{\eta}}}+
\frac{\tau}{4}F_{p_{\zeta}}e^{-\frac{\tau}{4} \frac{p_{\eta}}{2m_{\eta}}}
\left(\frac{\tau}{4}\frac{p_{\eta}}{2m_{\eta}}\right)^{-1}
\sinh\left[\frac{\tau}{4}\frac{p_{\eta}}{2m_{\eta}}\right]
\end{array}
\right\} :  U_B^{\rm BKNHC}\left(\frac{\tau}{4}\right)
$
\item
$
\left.
\begin{array}{ccl}
p & \to & p\exp\left[-\frac{\tau}{2} \frac{p_{\zeta}}{m_{\zeta}}\right]\\
q &\to & q\exp\left[-\frac{\tau}{2}\frac{p_{\xi}}{m_{\xi}}\right]\\
\zeta&\to &\zeta+\frac{\tau}{2} \frac{p_{\zeta}}{m_{\zeta}} \\
\xi &\to & \xi +\frac{\tau}{2} \frac{p_{\xi}}{m_{\xi}} \\
p_{\eta} & \to & p_{\eta} +\frac{\tau}{2}F_{p_{\eta}} \\
p_{\chi} &\to & p_{\chi} +\frac{\tau}{2}F_{p_{\chi}}
\end{array}
\right\} :  U_C^{\rm BKNHC}\left(\frac{\tau}{2}\right)
$
\item
$
\left.
\begin{array}{ccl}
q&\to &q+\frac{\tau}{4} \frac{p}{m} \\
\eta &\to & \eta + \frac{\tau}{4} \frac{p_{\eta}}{m_{\eta}} \\
p_{\zeta} & \to &
p_{\zeta}e^{-\frac{\tau}{4} \frac{p_{\eta}}{m_{\eta}}}+
\frac{\tau}{4}F_{p_{\zeta}}e^{-\frac{\tau}{4} \frac{p_{\eta}}{2m_{\eta}}}
\left(\frac{\tau}{4}\frac{p_{\eta}}{2m_{\eta}}\right)^{-1}
\sinh\left[\frac{\tau}{4}\frac{p_{\eta}}{2m_{\eta}}\right]
\end{array}
\right\} :  U_B^{\rm BKNHC}\left(\frac{\tau}{4}\right)
$
\end{itemize}


\newpage

\section{Numerical results}
\label{sec:num}

In its simplicity, the dynamics of a harmonic mode in one dimension
is a paradigmatic example for checking the chaotic (ergodic) properties
of constant-temperature phase space flows and the
correct sampling of the canonical distribution.
It is well known that it is necessary to generalize basic
Nos\'e-Hoover dynamics~\cite{Nose84,Hoover85,Nose91} to thermostats such as
the Nos\'e-Hoover chain~\cite{Martyna92,Martyna96} in order to produce
correct constant-temperature averages for systems such as the harmonic oscillator.

Some time ago, BK dynamics was devised
to provide a deterministic thermostat for systems such as classical spins \cite{Kusnezov90,Kusnezov92}.
To ensure efficient thermostatting, BK found it necessary to
introduce several `demons' per thermostatted degree of freedom,
where each demon was taken to have  a different and in principle complicated
coupling to the system degree of freedom \cite{Kusnezov90,Kusnezov92}.
In the present work, we keep the form of the system-thermostat coupling
as simple as possible, in order to facilitate the
formulation of explicit, reversible and measure-preserving integrators \cite{Ezra06}.
It is then of interest to investigate the ability of
our BK-type thermostats
to produce the correct canonical sampling
in the case of the harmonic oscillator.
Interest in harmonic modes is also justified
by the possibility of devising models of
condensed matter systems in terms of coupled
spins and harmonic modes, as already
done in quantum dynamics with so-called spin-boson models  \cite{Leggett87}.
We therefore investigate the performance of our
integration schemes on the simple one-dimensional harmonic oscillator.

For the particular calculations reported here, the oscillator
angular frequency, all masses and $k_B T$ were taken to be unity.
The time step in all cases was $\tau =0.0025$, and
all runs were calculated for $10^6$ time steps,
starting from the same initial conditions:
harmonic oscillator coordinate $q=0.3$,
all other phase space variables zero at $t=0$.

\begin{figure}[htb!]
\begin{center}
\includegraphics* {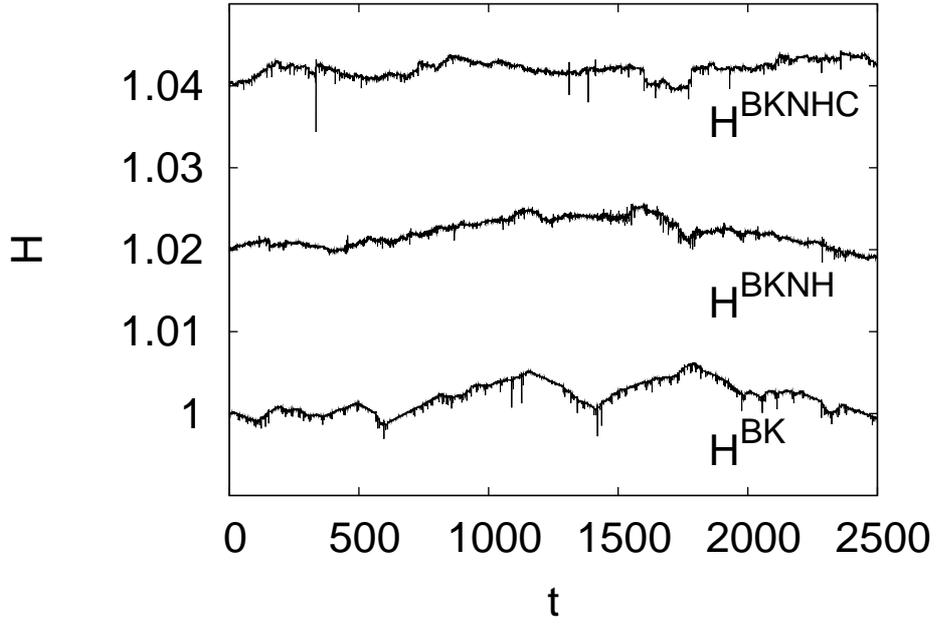}
\end{center}
\caption{
Comparison of the total extended Hamiltonian versus time
(normalized with respect to its value at $t=0$)
for the harmonic oscillator
undergoing simple Bulgac-Kusnezov dynamics ($H^{{\rm BK}}$),
NH controlled Bulgac-Kusnezov dynamics ($H^{{\rm BKNH}}$),
and Bulgac-Kusnezov-Nos\'e-Hoover chain dynamics ($H^{{\rm BKNHC}}$).
Two curves have been displaced vertically for clarity.
The time-reversible measure-preserving algorithms
developed in this paper conserve the extended Hamiltonian
to high accuracy in all three cases.
}
\label{fig:fig1}
\end{figure}

The measure-preserving algorithms derived here result in stable numerical
integration for all the three cases treated: BK, BKNH, and BKNHC chain dynamics.
Figure~\ref{fig:fig1} shows the three extended Hamiltonians (normalized by their respective
initial time value) versus time.
All three Hamiltonians are \emph{numerically} conserved by the corresponding
measure-preserving algorithm to very high accuracy (which is maintained
in all the three cases).

\begin{figure}[H]
\begin{center}
\includegraphics* {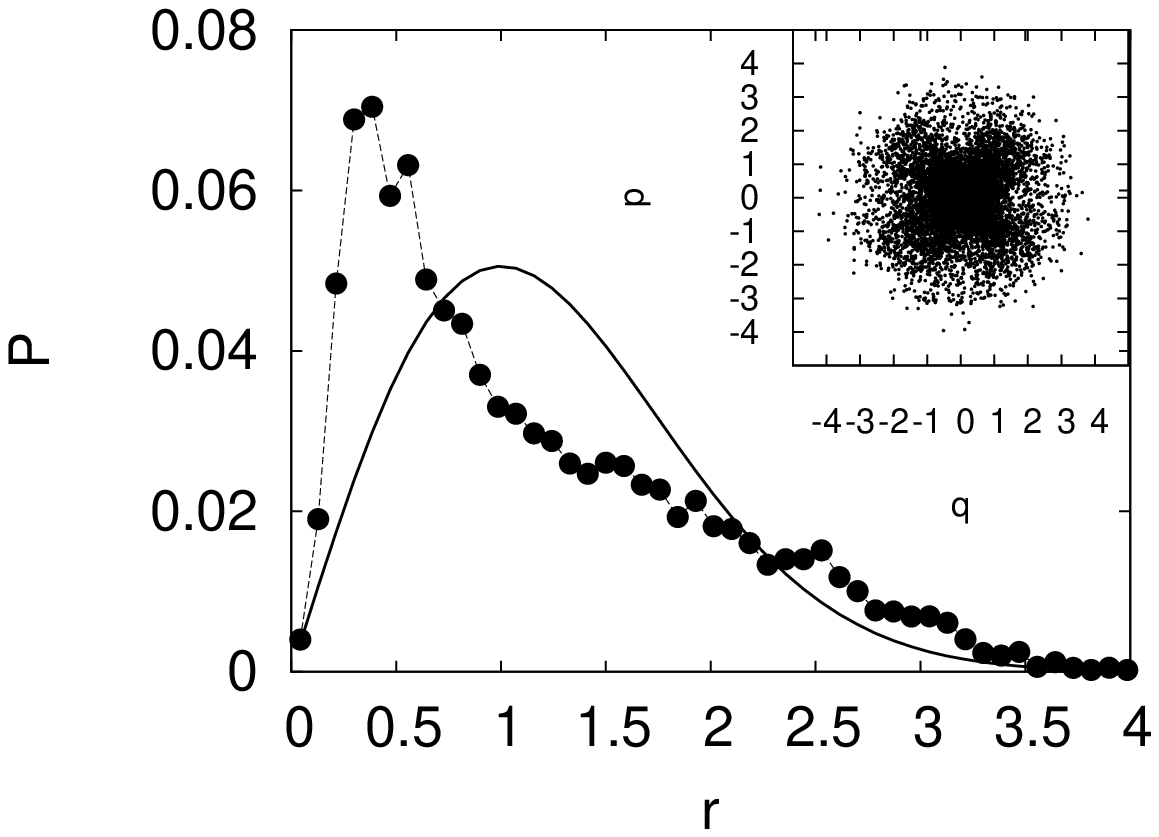}
\end{center}
\caption{
Radial phase space probability for a
harmonic oscillator under Bulgac-Kusnezov dynamics.
The continuous line
shows the theoretical value while the black bullets
display the numerical results.
The inset displays a plot of the phase space
distribution of points along the single trajectory
used to compute the radial probability.
}
\label{fig:fig2}
\end{figure}

However, the basic BK phase space flow is not capable of producing the correct canonical
sampling for a harmonic mode. This can be easily checked since
the canonical distribution function of the harmonic oscillator is isotropic
in phase space and its radial dependence can be calculated exactly.
Details of this way of visualizing the phase space sampling have already been
given in~\cite{Sergi01,Sergi03}.
Figure~\ref{fig:fig2}, displaying the comparison between the theoretical and
the calculated value of the radial probability in phase space, clearly shows that
the BK dynamics is not able to produce canonical sampling.
A look at the inset of Fig.~\ref{fig:fig2}, showing the phase space
distribution for  the harmonic mode, also immediately
shows that the dynamics is not ergodic.

\begin{figure}[H]
\begin{center}
\includegraphics* {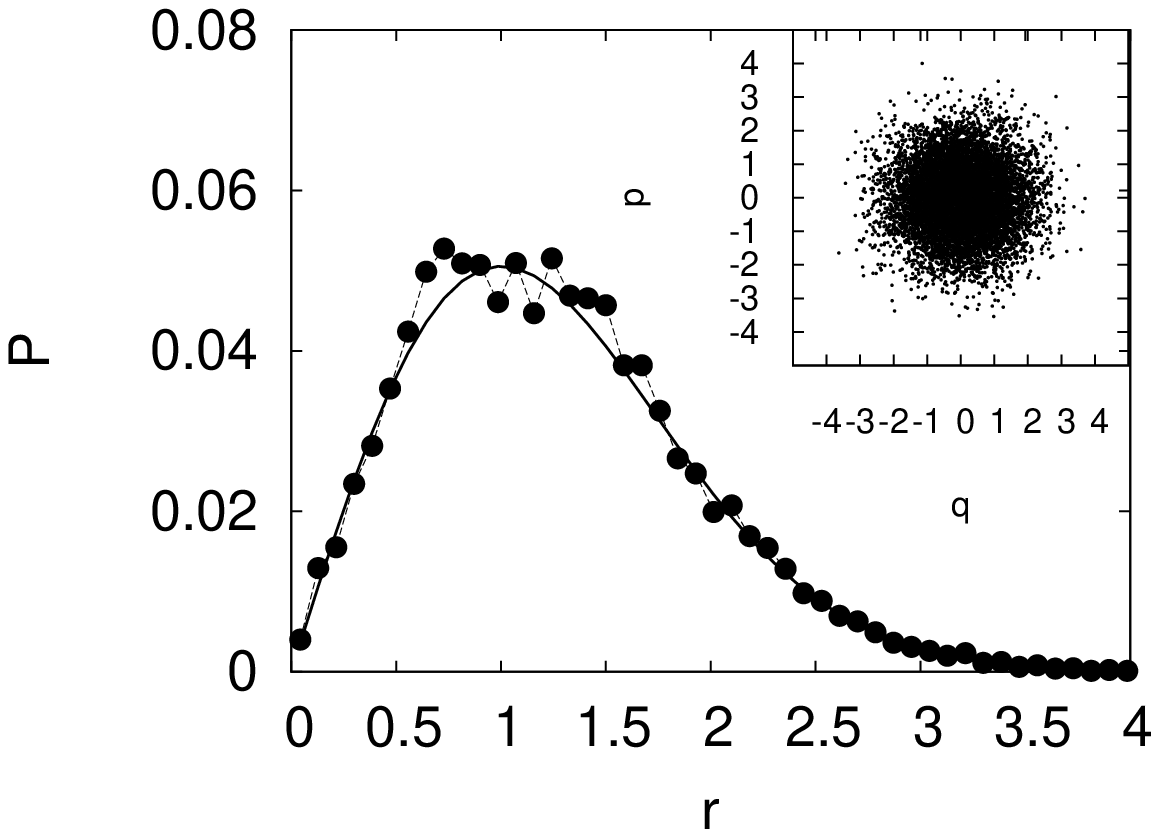}
\end{center}
\caption{
Radial phase space probability for
a harmonic oscillator under Nos\'e-Hoover controlled Bulgac-Kusnezov dynamics.
The continuous line
shows the theoretical value while the black bullets
display the numerical results.
The inset displays a plot of the phase space
distribution of points along the single trajectory
used to compute the radial probability.
}
\label{fig:fig3}
\end{figure}

\begin{figure}[H]
\begin{center}
\includegraphics* {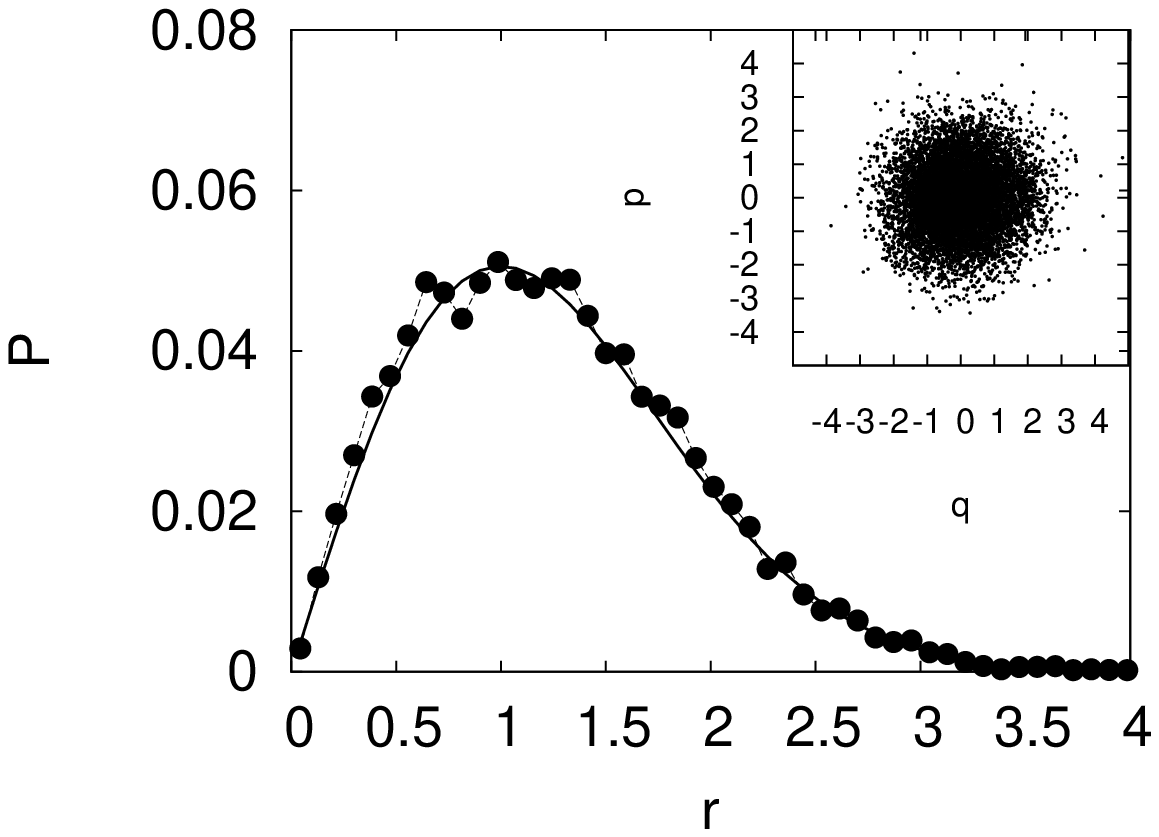}
\end{center}
\caption{
Radial phase space probability for
a harmonic oscillator under Bulgac-Kusnezov-Nos\'{e}-Hoover chain dynamics.
The continuous line
shows the theoretical value while the black bullets
display the numerical results.
The inset displays a plot of the phase space
distribution of points along the single trajectory
used to compute the radial probability.
}
\label{fig:fig4}
\end{figure}

The same analysis has been carried out for BKNH and BKNHC phase space flows,
and these are displayed in Fig~\ref{fig:fig3} and Fig~\ref{fig:fig4}, respectively.
Within numerical errors, both BKNH and BKNHC thermostats
are able to produce the correct canonical distribution
for the stiff harmonic modes.

Introduction of a single, global Nos\'{e}-type variable in the
BKNH thermostat effectively introduces additional coupling between the
two demon variables.  The effectiveness of the BKNH thermostat
is consistent with our findings (results not discussed here)
that introduction of explicit coupling between demons in BK
thermostat dynamics also leads to efficient thermostatting of
the harmonic oscillator.

\newpage

\section{Conclusions}
\label{sec:concl}

We have formulated Bulgac-Kusnezov \cite{Kusnezov90,Kusnezov92}, Nos\'e-Hoover
controlled Bulgac-Kusnezov, and Bulgac-Kusnezov-Nos\'e-Hoover chain
thermostats in phase space by means of non-Hamiltonian
brackets \cite{Sergi01,Sergi03}.
We have derived time-reversible measure-preserving algorithms \cite{Ezra06}
for these three cases and showed that additional control
by a single Nos\'e-Hoover thermostat or independent  Nos\'e-Hoover
thermostats is necessary to produce the correct canonical distribution
for a stiff harmonic mode.

Measure-preserving dynamics of the kind discussed here
is associated with equilibrium simulations 
(where, for example, there is a single temperature parameter $T$).
Stationary phase space distributions associated with non-equilibrium
situations are much more complicated than the smooth equilbrium
densities analyzed in the present paper \cite{Evans90,Hoover98b,Dorfman99}.
Nonequilibrium simulations of heat flow could be carried out 
by extending the present approach to multimode systems 
(e.g., a chain of oscillators) coupled to BK-type demons with associated NH
thermostats corresponding to two different temperatures \cite{Mundy00,Hoover07,Hoover07a}.

The techniques presented here for derivation and implementation
of thermostats have been shown to be efficient
and versatile.  We anticipate that
analogous approaches can be usefully applied to systems of classical spins
coupled to both harmonic and anharmonic modes.


\newpage

\appendix

\section{Operator formula}\label{app:app1}

We wish to determine the action of the propagator associated with the
Liouville operator Eq.\ \eqref{eq:operator_CBK}.
This is equivalent to solving the evolution equation (recall $i \neq k$)
\begin{equation}
\frac{d p_i}{d t} = \left(-\frac{p_k}{m_k}p_i+F_{p_i}\right)
\end{equation}
from $t =0$ to t=$\tau$.
Integrating, we  have
\begin{equation}
-\frac{m_k}{p_k}\left.\ln\left(-\frac{p_k}{m_k} p_i+F_{p_i}\right)\right\vert^\tau_0 = \tau
\end{equation}
giving
\begin{subequations}
\begin{align}
p_i(\tau) &\equiv \exp\left[\tau\left(-\frac{p_k}{m_k}p_i+F_{p_i}\right)\frac{\partial}{\partial p_i}\right] p_i\\
&= p_ie^{-\tau p_k/m_k}+\frac{m_k}{p_k}F_{p_i}\left(1-e^{-\tau p_k/m_k}\right) \\
&=
p_ie^{-\tau p_k/m_k}+
\tau F_{p_i}e^{-\tau \frac{p_k}{2m_k}}
\frac{\sinh\left[\tau\frac{p_k}{2m_k}\right]}
{\tau \frac{p_k}{2m_k}}\; .
\end{align}
\end{subequations}

\section{Invariant Measure of the BK phase space flows}
\label{app:BK-stat}

The phase space compressibility of the phase space BK thermostat
is
\begin{equation}
\kappa_{\rm BK}=\frac{\partial{\cal B}^{\rm BK}_{ij}}
{\partial x_i}\frac{\partial H_{\rm BK}}{\partial x_i}
=
-\frac{1}{m_{\zeta}}\frac{\partial G_1}{\partial p}\frac{\partial K_1}
{\partial p_{\zeta}}
-\frac{1}{m_{\xi}}\frac{\partial G_2}{\partial q}\frac{\partial K_2}
{\partial p_{\xi}}
\end{equation}
Upon introducing the function
\begin{equation}
H_{\rm T}^{\rm BK}
=H+\frac{K_1}{m_{\zeta}}+\frac{K_2}{m_{\xi}}
\end{equation}
one can easily find that
\begin{equation}
\kappa_{\rm BK}=\frac{1}{k_BT}\frac{dH_{\rm T}^{\rm BK}}{dt}
\end{equation}
so that the invariant measure in phase space reads
\begin{subequations}
\begin{align}
d\mu &= dx\,\exp\left[-\int_t dt\kappa_{\rm BK}\right] \\
& =
dx\,\exp\left[-\beta H_{\rm T}^{\rm BK}\right]\\
& = dx\, \exp[-\beta H^{\rm BK}]\exp[ \zeta + \xi]
\end{align}
\end{subequations}
as desired.

\section{Invariant Measure of the BKNH phase space flows}
\label{app:CBK-stat}

The phase space compressibility of the NH controlled Bulgac-Kusnezov thermostat
is
\begin{equation}
\kappa_{\rm BKNH}=\frac{\partial{\cal B}^{\rm BKNH}_{ij}}
{\partial x_i}\frac{\partial H_{\rm BKNH}}{\partial x_i}
=
-\frac{1}{m_{\zeta}}\frac{\partial G_1}{\partial p}\frac{\partial K_1}
{\partial p_{\zeta}}
-\frac{1}{m_{\xi}}\frac{\partial G_2}{\partial q}\frac{\partial K_2}
{\partial p_{\xi}}
-2\frac{p_{\eta}}{m_{\eta}}
\end{equation}
Upon introducing the function
\begin{equation}
H_{\rm T}^{\rm BKNH}
=H+\frac{K_1}{m_{\zeta}}+\frac{K_2}{m_{\xi}}
+\frac{p_{\eta}^2}{2m_{\eta}}
\end{equation}
we have
\begin{equation}
\kappa_{\rm BKNH}=\frac{1}{k_BT}\frac{dH_{\rm T}^{\rm BK}}{dt}
\end{equation}
so that the invariant measure in phase space is
\begin{subequations}
\begin{align}
d\mu &= dx\,\exp\left[-\int_t dt\kappa_{\rm BKNH}\right]\\
& =
dx\,\exp\left[-\beta H_{\rm T}^{\rm BKNH}\right] \\
& = dx\, \exp[-\beta H^{\rm BKNH}]\exp[\zeta + \xi + 2\eta].
\end{align}
\end{subequations}


\section{Invariant Measure of the BKNHC chain phase space flows}
\label{app:NBK-stat}

The phase space compressibility of the Nos\'e-Hoover-Bulgac-Kusnezov chain
is
\begin{equation}
\kappa_{\rm BKNHC}=\frac{\partial{\cal B}^{\rm BKNHC}_{ij}}
{\partial x_i}\frac{\partial H_{\rm BKNHC}}{\partial x_i}
=
-\frac{1}{m_{\zeta}}\frac{\partial G_1}{\partial p}\frac{\partial K_1}
{\partial p_{\zeta}}
-\frac{1}{m_{\xi}}\frac{\partial G_2}{\partial q}\frac{\partial K_2}
{\partial p_{\xi}}
-\frac{p_{\eta}}{m_{\eta}}-\frac{p_{\chi}}{m_{\chi}}
\end{equation}
Upon introducing the function
\begin{equation}
H_{\rm T}^{\rm BKNHC}
=H+\frac{K_1}{m_{\zeta}}+\frac{K_2}{m_{\xi}}
+\frac{p_{\eta}^2}{2m_{\eta}}+\frac{p_{\chi}^2}{2m_{\chi}}
\end{equation}
we have
\begin{equation}
\kappa_{\rm BKNHC}=\frac{1}{k_BT}\frac{dH_{\rm T}^{\rm BK}}{dt}
\end{equation}
so that the invariant measure in phase space reads
\begin{subequations}
\begin{align}
d\mu & =dx\exp\left[-\int_t dt\kappa_{\rm BKNHC}\right]\\
& =
dx\,\exp\left[-\beta H_{\rm T}^{\rm BKNHC}\right] \\
& = dx \, \exp[-\beta H^{\rm BKNHC}]\exp[\zeta + \xi + \eta +\chi].
\end{align}
\end{subequations}



\newpage


\end{document}